\documentclass[preprint,aps,nofootinbib,preprintnumbers,amsmath,amssymb]{revtex4}
\usepackage{graphicx}
\usepackage{dcolumn}
\usepackage{bm}
\usepackage{natbib}
\newcommand{\beq}{\begin{equation}}
\newcommand{\eq}{\end{equation}}
\newcommand{\bea}{\begin{eqnarray}\displaystyle}
\newcommand{\ea}{\end{eqnarray}}

\newcommand{\p}{\partial}

\newcommand{\nn}{\nonumber}

\newcommand{\half}{\frac{1}{2}}
\newcommand{\pb}{\partial_{\bar{z}}}
\newcommand{\pz}{\partial_z}
\newcommand{\zb}{\bar{z}}
\begin{document}
\preprint{Brown-HET-1494}
\preprint{MCTP 07-42}
\preprint{WITS-CTP-035}
\title{\Large\bf Matrix Model Maps and Reconstruction \\[6pt]
of AdS SUGRA Interactions }
\author{Sera Cremonini}
\email{seracre@umich.edu}
\affiliation{\it Michigan Center for Theoretical Physics, Randall
Laboratory of Physics,\\ The University of Michigan, Ann Arbor, MI
48109, USA }
\author{Robert de Mello Koch}
\email{robert@neo.phys.wits.ac.za}
\affiliation{\it Department of Physics, University of Witwatersrand,
Wits, 2050, South Africa }
\author{Antal Jevicki}
\email{antal@het.brown.edu}
\affiliation{\it Department of Physics, Brown University, Providence,
RI 02912, USA }
\date{\today}
\begin{abstract}
We consider the question of reconstructing (cubic) SUGRA interactions in AdS/CFT. The method we introduce
is based on the matrix model maps (MMP) which were previously successfully employed at the
linearized level.
The strategy is to start with the map for 1/2 BPS configurations which is exactly known
(to all orders) in the hamiltonian framework.
We then use the extension of the matrix model map with the corresponding Ward identities to
completely specify the interaction. A central point in this construction is the
non-vanishing of off-shell interactions (even for highest-weight states).
\end{abstract}
\maketitle
\section{Introduction}
The question of reconstructing bulk supergravity (SUGRA) through the AdS/CFT
\cite{Maldacena:1997re,Nastase:2007kj} correspondence is of considerable interest.
Initially  much insight into the correspondence was gained through the GKP-W holographic
relation which states \cite{Gubser:1998bc,Witten:1998qj} that correlators
of Yang-Mills theory coincide with certain boundary to boundary amplitudes in supergravity.
Indeed this was the scheme which provided some of the initial prescriptions for
relating cubic supergravity interactions to gauge theory correlators
\cite{Lee:1998bxa,Freedman:1998bj,Constable:2002hw,ArutyunovFrolov:1999,Mihailescu:1999,Berenstein:2002jq,Cremonini:2004kh,
Asano:2003xp,Dobashi:2002ar,Lee:2004cq}.
The holographic relation, however, has elements of an S-matrix relation,
and one can ask what set of correlators contains all the
information for reconstructing the theory in the bulk.
Although some studies \cite{Skenderis:2007yb} have been done along this direction,
there are still some main questions left open.
The issue/problem seems to be analogous to the question
of reconstructing the off-shell theory from strictly on-shell
data, a problem which is usually plagued by non-uniqueness. In addition
there is the question of unitarity, namely the issue of securing a unitary and local
evolution of the bulk theory.
An alternative is to develop the construction directly in the hamiltonian
framework, a method we consider in the present work.
The basic building block of our construction
will be the matrix model representation that was developed in the last few years
beginning with the 1/2 BPS sector of the theory.
This approach came for studies of 1/2 BPS correlators in gauge theory
\cite{Corley:2001zk,Berenstein:2004kk} and the dual
exact configurations in SUGRA \cite{Lin:2004nb,MaozRychkov:2005}.
What emerged is a fermion droplet correspondence (see \cite{Chen:2007du} and references therein).
Its hamiltonian version given through collective field theory
\cite{Jevicki:1979mb,deMelloKoch:2002nq}
can serve as a starting point for reconstructing the full theory.
Specifically, the strategy that we develop for the construction of the bulk
interaction is then as follows: starting from the non-zero
$c=1$ collective field theory vertex we proceed with the action
of raising operators to establish Ward identities that, as we
argue, are capable of determining the full cubic vertex.
The form of the raising and lowering operators can be  deduced through
the matrix model map (MMP) formulated in \cite{Donos:2005vm}.
The map of \cite{Donos:2005vm} was given at the linearized level, and was shown to
provide a mapping from  eigenfuctions of matrix model  equations
to those of AdS. As such our work represents an extension to the
nonlinear level of the mapping introduced in \cite{Donos:2005vm}.
The outline of this paper is as follows.
In Section \ref{SingleMatrixInt} we discuss the form of cubic interactions in supergravity
as well as for the 1/2 BPS  collective field theory.
Here we also discuss and resolve issues that concern the comparison of
the (vanishing) SUGRA vertex for the 1/2 BPS sector with
the (nonvanishing) matrix model vertex.
In Section \ref{SectMap} we review the linearized MMP of \cite{Donos:2005vm}
in terms of canonical transformations
on phase space.  This version turns out to be useful for the
nonlinear extension that we give in Section \ref{NonlinearAnalysis},
where we consider a simplified limit.
Finally, in Section \ref{FullVertex} we discuss the Ward identities and their ability to
determine the vertex (from the initial highest-weight one). Throughout this paper we
restrict our analysis to the $AdS_2$ case, where the method can be presented in the
simplest possible way.

\section{Cubic Interactions in the $1/2$ BPS  sector}
\label{SingleMatrixInt}

Our starting point is the direct hamiltonian level map that was
accomplished in the (limited) $1/2$ BPS sector of the theory.
On the Yang-Mills side one has a (reduced) matrix model hamiltonian
established in \cite{Corley:2001zk} and its collective field theory
hamiltonian description.
This is fully reproduced on the gravity side
with the 1/2 BPS reduction of 10D SUGRA in \cite{Lin:2004nb}.
In particular,
the energy of the 10D geometries of the $1/2$ BPS sector was
shown \cite{Lin:2004nb} to be given by
\beq
\label{LLMenergy}
E=\int dx_1 \int dx_2 \;(x_1^2+x_2^2)\; u(x_1,x_2) \, ,
\eq
where $u(x_1,x_2)$ is a density function distinguishing between
spacetime regions having different boundary conditions (``black'' and ``white'' regions).
The expression (\ref{LLMenergy}) is recognizable as the energy of fermions
(corresponding to matrix eigenvalues)
in a harmonic oscillator potential. In this language, $u$ is
responsible for differentiating
between particles (\emph{fermion droplets}) and holes.
After identifying $x_1=x$ and $x_2=y$, and after performing the $x_2$
integration over a ``black" region (fermion droplet), the energy (\ref{LLMenergy})
can be shown to be equivalent to the
collective field theory hamiltonian \cite{2D:1993}
\beq
\label{Hcoll}
H=\int dx \, \Bigl(\frac{y_+^3}{3}-\frac{y_-^3}{3}+x^2(y_+ - y_-)
\Bigr)\,
\eq
of a one-matrix model described by
\beq
\label{1DMM}
H=\half Tr(X^2+P^2)\, .
\eq
The Hermitian $N\times N$ matrix $X(t)$ depends only on time,
and $P(t)=\dot{X}(t)$ denotes its conjugate momentum.
The functions $y_+$ and $y_-$ describe the upper and
lower profiles of the Fermi droplet.
Furthermore, the matrix hamiltonian is related to (\ref{Hcoll})
via $X=x$ and $P=y$.
It is important to emphasize that the collective field formalism
describes well the \emph{fully interacting} theory of chiral primaries on
$AdS_5 \times S^5$.
To show this explicitly, we examine next the form of the cubic
vertex as given by collective field theory.
The dynamics of the resulting collective field theory can be
directly induced from the much simpler dynamics of the
one-dimensional matrix $X(t)$ (with eigenvalues $\lambda_i$),
after a change to the density field obeying  the following cubic collective hamiltonian:
\beq
H_{coll}= \int dx \, \Bigl( \, \half \, \p_x\Pi \; \phi \; \p_x\Pi +\frac{\pi^2}{6}\phi^3 +
 \half \, (x^2-\mu) \phi \, \Bigr).
\eq
The static ground state equation yields the background value $\phi_0$
for the field $\phi$.
One can then introduce small fluctuations about the
background, letting
\beq
\phi(x,t)=\phi_0(x)+\frac{1}{\sqrt{\pi}} \, \p_x \eta(x,t)\, .
\eq
After expanding the hamiltonian one finds
\beq
H=\int dx \, \Bigl[ \pi \phi_0 \Bigl(\half \Pi^2 +\half
(\p_x\eta)^2 \Bigr) + \frac{\pi^2}{6}(\p_x\eta)^3 + \frac{\pi}{2}
\Pi \, \p_x\eta \, \Pi \Bigr]\, .
\eq
Note that the corresponding quadratic Lagrangian takes the form
\beq
L_2 = \int dt \int dx \, \half \Bigl[ \frac{\dot{\eta}^2}{\pi \phi_0} - \pi \phi_0 \, \eta^2_{\, ,x}
\Bigr],
\eq
describing a massless particle in a gravitational background with
metric
\beq
g^0_{\mu\nu} = \bigl(\frac{1}{\pi \phi_0} , \pi \phi_0 \bigr).
\eq
The metric can be removed by an appropriate coordinate
transformation.
In terms of the ``time of flight'' coordinate $\tau$, the hamiltonian then becomes
\beq
H=\int d\tau \, \Bigl[\,  \half \, \Pi^2 +\half \,
(\p_\tau\eta)^2  + \frac{1}{6\pi^2 \phi_0^2}
\Bigl( (\p_\tau \eta)^3 + 3
\Pi \, \p_\tau \eta \, \Pi  \Bigr) \Bigr]\, .
\eq
Furthermore, notice that this is the theory of a massless boson with a spatially
dependent coupling, $g_{string}(\tau)=\frac{1}{\pi^2 \phi_0^2}.$
Since we are interested in studying the (cubic) interaction terms,
let's concentrate on
\beq
H^{(3)}=\int d\tau \, \frac{1}{6\pi^2 \phi_0^2}
\Bigl( (\p_\tau \eta)^3 + 3
\Pi \, \p_\tau \eta \, \Pi  \Bigr)\, .
\eq
If we recall how the (standard) fields $\alpha_\pm$ were introduced, $\alpha_\pm (x,t) =\p_x \Pi  \pm \pi
\phi(x,t)$, it is clear that they could have been expanded about
the background in a similar way:
\beq
\alpha_{\pm}=\pm \, \pi \phi_0 + \tilde{\alpha}_\pm.
\eq
The cubic hamiltonian takes on a much simpler form in terms of
$\tilde{\alpha}_\pm$:
\beq
H^{(3)}=\int_0^\pi \frac{d\tau}{\phi_0^2}\, \Bigl(\tilde{\alpha}_+^3(\tau) - \tilde{\alpha}_-^3(\tau)\Bigr).
\eq
This can be simplified even further by combining $\tilde{\alpha}_\pm$ into a
single field $\alpha(\tau)$ in the following way:
\bea
\alpha(\tau)&=&\tilde{\alpha}_+ (\tau) \;\;\;\text{for} \;\;\;\; \tau>0, \nn \\
&=&-\tilde{\alpha}_- (\tau) \;\;\;\text{for} \;\;\tau<0,
\ea
where we must now take $-\pi<\tau<\pi$.
Finally, expressing the cubic hamiltonian in terms of the new
field $\alpha$, we find
\beq
\label{H3alpha}
H^{(3)}=\int_{-\pi}^{\, \pi} \frac{d\tau}{\phi_0^2} \, \alpha^3(\tau).
\eq
Expanding $\alpha$ into creation and annihilation operators gives
\beq
\alpha(\tau) = \sum_n \sqrt{n} \bigl( e^{in\tau}a_n + e^{-in\tau}a_n^\dag
\bigr).
\eq
Rewriting (\ref{H3alpha}) in terms of creation and annihilation operators we
find several terms, but we would like to restrict our attention to
the one containing $a_1 a_2 a_3^\dag$ :
\bea
H^{(3)}&=&\sqrt{n_1\,n_2\,n_3} \int_{-\pi}^{\, \pi} \frac{d\tau}{\sin^2{\tau}} \,
e^{i(n_1+n_2-n_3)\tau} \; a_1 a_2 a_3^\dag + \dots \nn \\
&=& - \sqrt{n_1\,n_2\,n_3} \; \int \Bigl(\frac{d}{d\tau}\cot{\tau}\Bigr) \; e^{i(n_1+n_2-n_3)\tau} \; a_1 a_2 a_3^\dag + \dots \nn \\
&=& i (n_1+n_2-n_3) \; \sqrt{n_1\,n_2\,n_3} \; \int d\tau \cot{\tau}\, e^{i(n_1+n_2-n_3)\tau} \;
a_1 a_2 a_3^\dag + \dots \nn
\ea
where we have implicitly used the fact that the boundary term cancels.
Introducing $z=e^{i\tau}$ and letting $n\equiv n_1+n_2-n_3$, the integral above becomes
\beq
I=\int_{-\pi}^\pi d\tau \cot{\tau}\, e^{i \,n \,\tau}=
\int dz \frac{z^{n-1} }{(z-1)(z+1)}\, (z^2+1),
\eq
which has simple poles at $z_k=0,\pm 1$.
Evaluating the integral we find that the only non-zero
contribution from the residues occurs for $n>0$ and even, and is given by
$\sum_k \text{Res}(f,z_k)=2$, yielding
\beq
\label{CFTvertex}
H^{(3)} = -4\pi\sqrt{n_1 n_2 n_3}\;(n_1+n_2-n_3) \; a_1 a_2 a_3^\dag + \dots \; .
\eq
We should note that the vertex vanishes when $(n_1+n_2-n_3)=0$,
which is the on-shell energy conservation condition.
We mention here the relevance of an Euclidean picture which was established
recently in \cite{Jevicki:2006tr}.
It corresponds to the inverted harmonic oscillator model of the c=1 string theory
\cite{2D:1993}.
The analogue cubic hamiltonian interaction was shown capable of reconstructing
the non-critical string amplitudes at both tree and loop level. The relevance of this
S-matrix to AdS/CFT (and comparison with 1/2 BPS correlators)
was shown in \cite{Jevicki:2006tr}.
For completeness in the rest of this section we discuss the comparison of (\ref{CFTvertex})
with the SUGRA vertex obtained by studying three-point functions of chiral primaries
on $AdS_5 \times S^5$.
Next, we outline the main steps of such a comparison,
and leave a detailed discussion to Section \ref{details} below.
The typical 3-point (cubic) SUGRA interaction on
backgrounds of the form $AdS_n \times S^m$ is given by the overlap of bulk
wavefunctions,
\beq
H_3 = (\Delta_3-\Delta_1-\Delta_2)
\int_{AdS} d^{n-1}x \, \sqrt{g_{AdS}} \; g_{AdS}^{tt} \;
f_{I_1} f_{I_2} \bar{f}_{I_3}
\int_S d^my \, \sqrt{g_{S}} \; Y^{I_1} Y^{I_2} \bar{Y}^{I_3} \, ,
\eq
with $f_I(x)$ and $Y^I(y)$ denoting eigenfunctions on $AdS_n$ and $S^m$ respectively.
The total wavefunctions
$\psi(x,y)=\sum_I f_I(x) Y^I(y)$
obey the linearized equation
\beq
(\Box_{AdS}+\Box_S)\psi=0 \nn \, .
\eq
Understanding the cubic interaction then relies on understanding
bulk properties of $AdS$.
From the GKP-W map one has the ``holographic'' formula
\beq
H_3 \sim (\Delta_3-\Delta_1-\Delta_2) \, C(I_1,I_2,I_3) \, ,
\eq
where $C(I_1,I_2,I_3)$ are coefficients in the 3-point
correlator.
For the highest-weight states one has that their energy $\Delta$ is given by
the angular momenta $\Delta=j$.
One also has
\footnote{We note that in
Appendix \ref{appendix} we will show in more detail the origin
of the energy-conserving $\delta$-function in the 3-point function.}
\beq
C(j_1, j_2, j_3) \propto \delta_{j_1+j_2,j_3},
\eq
which is the  R-charge conservation condition.
We find that
the $\delta$-function forces the (highest-weight) vertex to vanish, $V^{\,h. w.}=0$.
We emphasize that this
implies that the holographic vertex is equal to 0 both on and off- shell.
On the other hand, the collective vertex  is seen to be  \emph{non-vanishing}  off-shell
and can therefore serve as the starting point for a raising-lowering procedure that  one
can apply to highest-weight states.
\subsection{Chiral Primary Interactions in $AdS_5 \times S^5$}
\label{details}
Let's now examine the full interacting theory of chiral primaries, with the
ultimate goal of showing agreement with the collective field calculation.
We consider the case of $AdS_5 \times S^5$, which has been studied in \cite{Lee:1998bxa}.
The equation of motion for the chiral primary field $s$, of mass
$m^2=j(j-4)$, was found to be of the form
\beq
(\nabla_\mu \nabla^\mu-m_I^2)s^I=
\sum_{J,K} \Bigl( D_{IJK} s^J s^K + E_{IJK} \nabla_\mu s^J \nabla^\mu s^K
+F_{IJK} \nabla^{(\mu} \nabla^{\nu )} s^J \nabla_{(\mu} \nabla_{\nu )}
s^K\Bigr), \nn
\eq
where $\mu$ denotes $AdS_5$ coordinates, and the sphere dependence
has already been integrated out.
For the explicit form of the coefficients $D,E$ and $F$ we refer
the reader to \cite{Lee:1998bxa}.
The derivative terms can be removed by the following field redefinition
\beq
s^{\,I}=s^{\,\prime I} + \sum_{J,K} \Bigl(J_{IJK}s^{\,\prime J} s^{\, \prime K} +
L_{IJK} \nabla^\mu  s^{\, \prime J}  \nabla_\mu s^{\, \prime K}
\Bigr) \, ,
\eq
where $L_{IJK}=\half F_{IJK}$ and  $J_{IJK}=\half E_{IJK} +\frac{1}{4}
F_{IJK} (m_I^2-m_J^2-m_K^2+8)$.
The field redefinition dramatically simplifies the equation of motion, which becomes
\beq
(\nabla_\mu \nabla^\mu-m_I^2)s^I= \sum_{J,K} \lambda_{IJK} \, s^J s^K \, ,
\eq
where $\lambda_{IJK}=D_{IJK}-(m_J^2+m_K^2-m_I^2)
J_{IJK}-\frac{2}{5}L_{IJK} \, m_J^2 \, m_K^2$.
Finally, after plugging in the coefficients $D,E$ and $F$, the action for the
chiral primary $s$ becomes
\beq
\label{ads5Action}
S = \int d^5x \sqrt{-g} \;\Bigl[ -\nabla s^I \,\nabla \bar{s}^I  - m_I^2
|s^I|^2-\half \lambda_{IJK} \bigl(s^I s^J \bar{s}^K+ c.c \bigr) \Bigr]\, ,
\eq
where $m^2=j(j-4)$, and the coupling constant \cite{Lee:1998bxa} is
\bea
\lambda_{123} &=& (j_3-j_1-j_2)\; 2 \kappa
\frac{\sqrt{j_1 j_2 j_3 (j_3^2-1)(j_3+2)}(j_3-2)}
{\sqrt{(j_1^2-1)(j_2^2-1)(j_1+2)(j_2+2)}} \times f_{123} \, , \nn \\
f_{123} &=& \frac{1}{\sqrt{2\pi^3}}
\frac{\sqrt{(j_1+1)(j_1+2) (j_2+1) (j_2+2)}}
{\sqrt{(j_3+1)(j_3+2)}}\, .
\ea
The coefficient $f_{123}$ comes from the overlap integral over
spherical harmonics on $S^5$.
In global coordinates, the highest-weight state on $AdS_5 \times S^5$ takes the form
\beq
s=\frac{\sqrt{\Delta(\Delta-1)}}{\pi (\cosh{\mu})^\Delta}\, .
\eq
The matrix element of the cubic hamiltonian for the action
(\ref{ads5Action}) is then given by
\bea
<3|H_3|12> &=& \frac{1}{2^{3/2}\pi} \frac{(\Delta_1-1)(\Delta_2-1)(\Delta_3-1)}
{(\Delta_3-1)(\Delta_3-2)} \times G_{123}\; \delta(j_1+j_2-j_3)\nn \\
&=& (\Delta_3-\Delta_1-\Delta_2)
\frac{\sqrt{\Delta_1\Delta_2\Delta_3}}{N}\; \delta(j_1+j_2-j_3)\, .
\ea
Since $\Delta=j$, this agrees with the collective field theory
vertex (\ref{CFTvertex})
\beq
H^{(3)} \sim -4\pi\sqrt{n_1 n_2 n_3}\;(n_1+n_2-n_3) \; a_1 a_2
a_3^\dag,
\eq
apart from the absence of the $\delta$-function coming from
conservation of angular momentum.
Thus, we have shown that the collective field theory
vertex is contained in the gravity description. We will discuss
at a later stage the origin of two different pictures (for 1/2 BPS states)
related to the appearance of a delta function term in the vertex.

\section{Matrix Model Maps}
\label{SectMap}

Our goal is to extend the hamiltonian formulation from the highest-weight states
of the bubbling 1/2 BPS configuration.
In global coordinates $AdS_5 \times S^5$ can be written as
\beq
ds^2 = (-\cosh^2\rho dt^2 + d\rho^2 + \sinh^2\rho d\Omega_3^2)+
(\sin^2\theta d\theta^2 + d\phi^2 + \cos^2\theta
d\tilde{\Omega}_3^2),
\eq
and, as we will show in more detail later, the chiral primary fields fluctuations read
\beq
\delta g \sim \Bigl( \frac{\cos\theta}{\cosh\rho} e^{i \phi} \Bigr)^j \, ,
\eq
a highest-weight state of the isometry algebra.
The collective droplet vertex represents an off-shell interaction of such fluctuations.
The basic strategy that we will employ is then
to use the resulting non-vanishing three-point interaction as a starting point for
reconstructing the full 3-point vertex,
\emph{i.e.} involving more general states.
The first ingredient in this program is the reconstruction of linearized wavefunctions:
\beq
\psi_{jnm}(t,\rho,\theta,\phi)\sim L_+^n J_-^{j-m}
\psi_j\Bigl(\frac{\cos\theta}{\cosh\rho} e^{i \phi}\Bigr) \, .
\eq
This was done in \cite{Donos:2005vm}.
From the interactions of chiral primaries, we will develop an analogous
raising-lowering Ward identity which relates
$V_{j_1 n_1 \, j_2 n_2 \, j_3 n_3}$, the vertex for two-matrix states, to $V_{j_1 j_2 j_3}$,
the one-matrix vertex.
Toward this end it is important to describe the inclusion of  the $1/2$ BPS
``bubbling'' configurations of $AdS_5 \times S^5$ in the two matrix (coordinate) picture.
In the 2D coordinate space $(X_1,X_2)$, where the hamiltonian is given by
$H=\half(X_1^2+X_2^2+P_1^2+P_2^2)$ and the angular momentum by
$J=X_1 P_2 - X_2 P_1$,
one can introduce complex coordinates
\beq
Z=\frac{X_1+iX_2}{\sqrt{2}}, \;\;\;\; \bar{Z}=\frac{X_1-iX_2}{\sqrt{2}} \, ,
\eq
with corresponding conjugate momenta
\beq
\Pi=\frac{P_1+iP_2}{\sqrt{2}}, \;\;\;\; \bar{\Pi}=\frac{P_1-iP_2}{\sqrt{2}} \,.
\eq
Switching to creation and annihilation operators,
\bea
Z&=&\frac{1}{\sqrt{2}}(A^\dag+B), \;\;\;\; \bar{Z}=
\frac{1}{\sqrt{2}}(A+B^\dag) \, , \\
\Pi&=&\frac{-i}{\sqrt{2}}(A^\dag-B), \;\;\;\;
\bar{\Pi}=\frac{i}{\sqrt{2}}(A-B^\dag)\, ,
\ea
the hamiltonian and angular momentum generators become
\bea
H &=& Tr(A^\dag A + B^\dag B) \, , \nn \\
J &=& Tr(A^\dag A - B^\dag B) \, .
\ea
So $1/2$ $BPS$ states having $H=J$ are described by a truncation to the sector where
$B=0$, and only $A$ oscillators remain. This condition can be translated into having a
single matrix $X=(A+A^\dag)$, with conjugate momentum
$P=i(A-A^\dag)$.
In the phase space (matrix model) one has the corresponding canonical
transformation
\bea
X &=& \frac{X_1+P_2}{\sqrt{2}}, \;\;\;\; Y=\frac{X_1-P_2}{\sqrt{2}},
\nn \\
P_X &=& \frac{P_1-X_2}{\sqrt{2}}, \;\;\;\; P_Y=\frac{P_1+X_2}{\sqrt{2}},
\ea
with the fact that
\beq
J=X_1 P_2-X_2 P_1=\half (X^2+P_X^2)-\half (Y^2+P_Y^2)=\tilde{J}.
\eq
In the matrix model picture (which from now on we denote by a \emph{tilde}),
the R-charge transformation is not a coordinate transformation, but rather a canonical
transformation (dynamical symmetry). This gives an explanation of the origin of the
non-conservation in the $3$-vertex of the cubic collective field theory that we have noted
earlier: in the matrix model formulation we have two
representations that are related by a canonical transformation.
Next, we describe the matrix model map associated with
LLM (\emph{one-matrix} model), followed by the construction of \cite{Donos:2005vm}
which extends it to \emph{two matrices}.
The LLM map is given by one central formula
\beq
Z(x_1,x_2,y)= \frac{y^2}{\pi} \int_D dx_1^\prime dx_2^\prime \; \frac{ u(x_1^\prime,x_2^\prime,0) }
{[(\vec{x}-\vec{x}^{\, \prime})^2+y^2]^2} \, ,
\eq
where the integral is defined over a domain $D$ and $u(x_1,x_2,0)=\pm\half$.
It is a non-linear map since the dynamical degree of freedom on the
right hand side is not $u(x_1,x_2,0)$, but the boundary of the
domain.
Linearization leads to the following (linear) relationship
(for a detailed derivation see \cite{Donos:2005vm}):
\beq
\label{LLMmap}
\delta g = \frac{1}{2\pi} \, \int_0^{2\pi} d\tau \; \frac{1-4a^2-a^4+4a^3
\cos(\tau-\phi)}{[1+a^2-2a\cos(\tau-\phi)]^2}\; \delta
\alpha(\tau) \, ,
\eq
where $a=\frac{\cos\theta}{\cosh\rho}$.
On the right-hand side of the equation we have the small fluctuation $\delta \alpha(\tau)$ of the
one-matrix collective field described by
\beq
H_2 = \half \int dx \, \pi \phi_0(x) \, \Bigl(\Pi^2 + (\p_x \eta)^2\Bigr) = \int
d\tau (\delta \alpha(\tau))^2 \, ,
\eq
with
\beq
\phi_0(x)=\frac{1}{\pi} \sqrt{\mu-x^2} \, ,
\eq
and $d\tau = \frac{dx}{\pi \phi_0(x)}$.
On the left-hand side of (\ref{LLMmap}) $\delta g$ denotes the fluctuation of a
SUGRA chiral primary.
In the notation of \cite{Donos:2005vm}, $t$ and $\rho$ denote $AdS$ coordinates,
while $\theta$ and $\phi$ are sphere angles.
For $\delta \alpha(\tau)\sim e^{ij\tau}$ one gets
\beq
\delta g \sim \Bigl(\frac{\cos\theta}{\cosh\rho}e^{i\phi}\Bigr)^j \, ,
\eq
the correct chiral primary fields fluctuations.
More precisely, denoting the Kernel by $K_{LLM}(\rho,\theta,\phi;\tau)$, one finds
(see \cite{Donos:2005vm} for more details):
\beq
\delta g (t,\rho,\theta,\phi) =\frac{e^{\,ijt}}{2\pi} \int_0^{2\pi} d\tau \,
K_{LLM}(\rho,\theta,\phi;\tau)\, \delta\alpha(\tau) \, .
\eq
This is a \emph{one-dimensional} map from the space
$\tau = \int \frac{dx}{\pi \, \phi_0(x)}$ of a matrix model
to the subspace of $AdS_5 \times S^5$
given by $\frac{\cos\theta}{\cosh\rho}\, e^{i\phi}$.
The extension of the linearized LLM map to the two-matrix case
was given in \cite{Donos:2005vm} and starts with the
matrix observable
\beq
\psi(x,n)=Tr\Bigl(\bigl(\delta(x-(A+A^\dag))B^n \bigr)_{SYM}
\Bigr)\, .
\eq
This then leads to an eigenvalue problem
\beq
\hat{K}\psi=\omega \, \psi \, ,
\eq
with solution
\beq
\tilde{\psi}_{jn}(\tau,\sigma) = \sin\bigl((j+2n)\tau\bigr)\,e^{in\sigma},
\;\;\;\;\; \omega_{jn}=j+2n \, .
\eq
Through a kernel constructed in \cite{Donos:2005vm}, this maps into
a non-trivial eigenfunction on $AdS$ space:
\beq
\psi_{jn}(t,\rho,\phi, \theta)= e^{i \omega_{jn}t}\, \cos^j\theta \, \frac{1}{4\pi^2}\int_0^{2\pi}d\tau
\int_0^{2\pi} d\sigma \, K_2(\rho,\phi;\tau,\sigma)\, \tilde{\psi}_{jn} \, .
\eq
Notice that the map is $2 \leftrightarrow 2$, mapping the two coordinates
$\tau, \sigma$ of the matrix model into the spacetime coordinates $\rho, \phi$.
Furthermore,
we have the following two remarks about the kernel $K_2$.
First, when applied to the states with $n=0$, it reduces to the
kernel associated with the LLM map.
Second, the map is essentially a reduction to action-angle variables
associated with the non-trivial $AdS$ Laplacian.

\section{Non-linear Analysis}
\label{NonlinearAnalysis}

We now come to the main consideration of this work and address the
question of a non-linear extension. In this section we will also address the issue
concerning the presence of delta function constraint in the 1/2 BPS interaction vertex. To simplify the discussion we
start by considering what we refer to as the \emph{non-relativistic model},
which will allow us to present the main steps of our argument in explicit terms.
Recall that in Section \ref{SectMap} we distinguished between the
matrix model picture (\emph{i.e.} the tilde representation with matrices $X,Y$ and
conjugate momenta $P_X,P_Y$)
and the coordinate space ($X_1,X_2$).
In the non-relativistic approximation one directly replace the matrices
with the corresponding coordinates, a procedure that is simple to
implement based on density fields. In Section III we described (at the matrix level)
the canonical transformation relating the two pictures in question, with
\beq
\tilde{H} = \half(x^2+y^2+p_x^2+p_y^2)\, ,
\eq
and similarly for $H$.
The linear map (in the non-relativistic approximation) which relates the
two representations reads
\beq
\tilde{\psi}(x,y)=\int dx_1 \int dx_2 \, K(\tilde{x},\vec{x})\;
\psi(x_1,x_2)\, ,
\eq
where the kernel is given by
\beq
\label{BasisChange}
K(x,y;x_1,x_2)=\frac{1}{\sqrt{2\pi}} \; e^{-ix_2\frac{x-y}{\sqrt{2}}}
\, \delta(x_1-\frac{x+y}{\sqrt{2}})\, .
\eq
It corresponds to a canonical transformation such that
\beq
J=x_1p_2-x_2p_1=\half(p_x^2+x^2)-\half(p_y^2+y^2)=\tilde{J} ,
\eq
related to the change to the matrix model picture discussed in Section \ref{SectMap}.
It maps (matrix model) eigenfunctions
\beq
\label{HerWavefn}
\tilde{\psi}_{jn}(x,y)=e^{-(x^2+y^2)}H_{j+n}(x)H_n(y)
\eq
into (space-time) eigenfunctions
\beq
\label{LagWavefn}
\psi_{jn}(r,\phi)=\frac{e^{ij\phi}}{\sqrt{2\pi}}L_n^j(r)\, ,
\eq
where in the space-time picture $x_1+ix_2=re^{i\phi}$.
The vertices in the two pictures are denoted by $V$ and
$\tilde{V}$ and are given by the overlap integral of three
eigenfunctions:
\beq
V_{j_1n_1 j_2 n_2 j_3 n_3}=(\Delta_1-\Delta_2-\Delta_3)\int
\frac{d^2\vec{x}}{\sqrt{\phi_0(x)}} \,
\psi_{j_1n_1}^\ast(r,\phi) \, \psi_{j_2n_2}(r,\phi) \, \psi_{j_3n_3}(r,\phi) \, ,
\eq
with $\psi_{j n}$ given in (\ref{LagWavefn}),
and similarly for $\tilde{V}$.
The 3-point overlap $V_{j_1n_1 j_2 n_2 j_3 n_3}$ will then be roughly of the form
\bea
V_3 &\sim& \int d\phi \; e^{i(-j_1+j_2+j_3)\phi} \; \int dr \, L_{n_1}^{j_1} \,
L_{n_2}^{j_2}\, L_{n_3}^{j_3} \nn \\
&\sim& \delta(-j_1+j_2+j_3) \int dr \, L_{n_1}^{j_1} \,
L_{n_2}^{j_2}\, L_{n_3}^{j_3} \; \equiv \; \delta(-j_1+j_2+j_3) \; \mathcal{V} \, ,
\ea
and (still) yield a conserving delta function.
Let's briefly sketch what happens in the case of the
\emph{tilde} representation, with eigenfunctions now given by
(\ref{HerWavefn}).
The overlap integral takes the form
\beq
\tilde{V}_3 \sim \int dx  \; e^{-3x^2} H_{j_1+n_1}(x) \, H_{j_2+n_2}(x) \,
H_{j_3+n_3}(x)
\int dy \; e^{-3y^2}H_{n_1}(y)\, H_{n_2}(y) \,H_{n_3}(y)\, .
\eq
As one can verify, written in this basis the vertex no longer has a conserving
$\delta$-function.
Thus, as we commented earlier, the vertex $V$ has R-charge
conservation
\beq
V_{j_1n_1 j_2 n_2 j_3 n_3}=\delta_{j_1,j_2+j_3} {\mathcal V},
\eq
while $\tilde{V}$ does not.
This is explained by the different action of the R-charge operator
$\hat{J}$ in the two pictures.
While in the present case one can easily show that for 2-point
overlaps
\beq
\int dx_1 dx_2 \, \psi_{jn} \psi_{j^\prime n^\prime} =
\int dx dy \, \tilde{\psi}_{jn} \tilde{\psi}_{j^\prime n^\prime} \, ,
\eq
one cannot do that for the 3-point function overlap.
In fact one can show explicitly that
\beq
V_{j_1n_1 j_2 n_2 j_3 n_3} \neq \tilde{V}_{j_1n_1 j_2 n_2 j_3 n_3} \, .
\eq
The basic theorem that we will establish in what follows is that
the two hamiltonians
\beq
\label{hnontilde}
H=\omega_{jn} A^\dag_{jn}A_{jn} + \bigl(V_{j_1n_1 j_2 n_2 j_3 n_3}A^\dag_{j_1 n_1}
A_{j_2 n_2}A_{j_3 n_3} + h.c.\bigr)
\eq
and
\bea
\label{htilde1}
\tilde{H} &=& \sum_{j\,n} \omega_{jn} \tilde{A}^\dag_{jn}\tilde{A}_{jn} + \nn \\
&+& \sum_{\{j's, \, n's\}}
\Bigl( \tilde{V}^{(1)}_{j_1n_1 j_2 n_2 j_3 n_3} \, \tilde{A}^\dag_{j_1 n_1} \tilde{A}_{j_2 n_2}
\tilde{A}_{j_3 n_3} +\tilde{V}^{(2)}_{j_1n_1 j_2 n_2 j_3 n_3}
\tilde{A}^\dag_{j_1n_1} \tilde{A}^\dag_{j_2 n_2}\tilde{A}^\dag_{j_3 n_3}
+ h.c. \Bigr)
\ea
are in fact equivalent, with a  non-linear canonical  transformation relating them.
To demonstrate this  statement, namely the fact  that (\ref{hnontilde}) and (\ref{htilde1})
match,
we would like to perform the following field redefinition:
\beq
\label{fieldred}
\tilde{A}_N=A_N+c_{N M P}A_M A_P + d_{N M P}A^\dag_M A_P + e_{N M P}A^\dag_M
A^\dag_P \, .
\eq
We have simplified the notation by using the index $N$ to denote all quantum
numbers $(j,n)$.
The hamiltonian in the \emph{tilde} representation with this more compact
notation takes the form
\bea
\label{htilde2}
\tilde{H}&=& \tilde{H}_2 +\tilde{H}_3 \nn \\
&=& \sum_N \omega_{N} \tilde{A}^\dag_{N}\tilde{A}_{N} +
\sum_{\{N,M,P\}}
\Bigl( \tilde{V}^{(1)}_{NMP} \, \tilde{A}^\dag_{N} \tilde{A}_{M}\tilde{A}_{P}
+\tilde{V}^{(2)}_{NMP}
\tilde{A}^\dag_{N} \tilde{A}^\dag_{M}\tilde{A}^\dag_{P}+ h.c. \Bigr)\, .
\ea
Under the field redefinition (\ref{fieldred}) the quadratic part $\tilde{H}_2$
yields additional cubic terms, and
the total hamiltonian becomes
\bea
\tilde{H}
&=& \sum_{N} \omega_{N} A^\dag_{N} A_{N} + \sum_{N,M,P}
\Bigl[ \bigl(\tilde{V}^{(1)}_{N M P}+\omega_{N}c_{N M P}+\omega_{P}\bar{d}_{P M
N}\bigr) A_N^\dag A_M A_P + \nn \\
&+& (\tilde{V}^{(2)}_{NMP}+\omega_N e_{NMP})A_N^\dag A^\dag_M A^\dag_P + h.c.\bigr]
\ea
If we want this to match (\ref{hnontilde}), we need the following conditions
on the coefficients of the field redefinition:
\beq
\omega_N \, e_{NMP}=-\tilde{V}^{(2)}_{NMP}, \;\;\;
\tilde{V}^{(1)}_{NMP}+\omega_{N} \, c_{N M P}+\omega_{P}\, \bar{d}_{P M N}=V_{NMP} \, .
\eq
Furthermore, we can obtain additional constraints on
$c_{NMP}$, $d_{NMP}$ and $e_{NMP}$ by imposing appropriate commutation
relations:
\bea
\label{comm1}
[\tilde{A}_N,\tilde{A}_{N'}]&=& 0 \, , \\
\label{comm2}
[\tilde{A}_N,\tilde{A}^\dag_{N'}]&=&\delta_{N,N'} \, .
\ea
Requiring (\ref{comm1}) yields
\beq
d_{N' N M} = d_{N N' M} \, , \;\;\;\;
e_{N' M N} -e_{N M N'} + e_{N' N M} - e_{N N' M} =0 \, .
\eq
The remaining commutation relation (\ref{comm2}) gives
\bea
d_{N M N'} + \bar{c}_{N' M N} + \bar{c}_{N' N M} &=& 0 \, , \nn \\
\bar{d}_{N' M N} + c_{N M N'} + c_{N N' M} &=& 0 \, .
\ea
This entirely fixes the non-linear redefinition (\ref{fieldred}),
showing that one can in fact connect the two hamiltonians.

To summarize,
we have described how the matrix level canonical transformation induces changes at the
nonlinear level. One has two related pictures, one in which the R-symmetry
is implemented as a coordinate symmetry (with the corresponding delta function) and
another where the R-symmetry is dynamical, given
by a canonical transformation.
We have shown the equivalence of these two pictures through a
nonlinear field transformation.
Related field transformations have been identified previously at the Lagrangian level
in \cite{ArutyunovFrolov:1999}.

\section{ Ward Identities and Vertex Reconstruction}
\label{FullVertex}

Our main goal is to establish that, starting from the vertex of
highest-weight states, it is possible to build the vertex for more general
states that are reachable by (in this case) $SL(2)$ raising/lowering procedure.
Specifically, we will develop an identity that will
allow us to generate such non-trivial vertices, by making use of
the available Ward identities.
We will again start from the simplified
(non-relativistic) model discussed in Section \ref{NonlinearAnalysis}.
This will then be followed
by a discussion on the form of Ward identities in the AdS case.
\subsection{Non-relativistic Model}
Recall that the hamiltonian of the non-relativistic model
is given by
\beq
H=\frac{x^2+y^2}{2}+\frac{p_x^2+p_y^2}{2}\, ,
\eq
or, in terms of creation and annihilation operators,
\beq
H=a^\dag a + b^\dag b \, .
\eq
Let's introduce complex variables
\beq
z=\frac{a^\dag+b}{\sqrt{2}}\, , \;\;\;\; \bar{z}=\frac{a+b^\dag}{\sqrt{2}}\, ,
\eq
and corresponding conjugate momenta
\beq
\Pi=-i\pb=i\frac{a^\dag-b}{\sqrt{2}}\, , \;\;\;\;
\bar{\Pi}=-i\pz=-i\frac{a-b^\dag}{\sqrt{2}}\, .
\eq
These expressions can be combined to obtain
\beq
a=\frac{1}{\sqrt{2}}(\bar{z}+\partial_z)\, , \;\;\;\;\;
b=\frac{1}{\sqrt{2}}(z+\partial_{\bar{z}})\, .
\eq
The wavefunctions are then given by
\beq
|J,n>\, \equiv \frac{(a^\dag)^{J+n} \,(b^\dag)^{n}}{\sqrt{(J+n)!\, n!}}\, |0> \, ,
\eq
and the generators $l_+,l_-,l_0$ by
\bea
\label{lp}
l_+ &=&
\frac{1}{2}(\partial_z\partial_{\bar{z}}+z\bar{z}-1)-\frac{z}{2}\partial_z-\frac{\bar{z}}{2}\partial_{\bar{z}}\, , \\
l_- &=&
\frac{1}{2}(\partial_z\partial_{\bar{z}}+z\bar{z}+1)+\frac{z}{2}\partial_z+\frac{\bar{z}}{2}\partial_{\bar{z}}\, , \\
l_0 &=& -\partial_z\partial_{\bar{z}}+z\bar{z}-1 \, .
\ea
As we mentioned earlier, our goal is to use these generators to derive an
identity for the cubic vertex, which is given by
\beq
\int d^2 x \frac{1}{\sqrt{\phi_0(\vec{x})}} \; \eta \,\eta \, l_0
\eta \, ,
\eq
with $\vec{x}=(x,y)$.
If we plug the standard mode expansion
\beq
\eta=\sqrt{2} \sum_{J=1}^\infty \sum_{n=0}^\infty
(\bar{c}_{J,n}\psi_{J,n}+c_{J,n}\bar{\psi}_{J,n})
\eq
into the vertex we find
\bea
\sum_{J=1} \sum_{n=0}\; \Bigl[\bar{c}_{J_1,n_1}\bar{c}_{J_2,n_2}c_{J_3,n_3}
\int dx\int dy \frac{1}{\sqrt{\phi_0(\vec{x})}} \; \psi_{J_1,n_1}
\psi_{J_2,n_2}\bar{\psi}_{J_3,n_3}+\ldots\Bigr]\,P \, ,
\ea
where we denote by $P$ the prefactor coming from the action of
$l_0$ on the wavefunctions.
Next, we would like to use the fact that
\beq
\psi_{J,n}=\frac{l_+}{\sqrt{(J+n)n}}\, \psi_{J,n-1}\, ,
\eq
and focus on
\beq
V_1 = \int d^2 x \frac{1}{\sqrt{\phi_0(\vec{x})}} \, \bar{\psi}_{J_1,n_1}
\psi_{J_2,n_2} \psi_{J_3,n_3} \, .
\eq
Using (\ref{lp}), the vertex term above becomes
\beq
V_1 = \int d^2 x \frac{1}{2\sqrt{\phi_0(\vec{x})}\sqrt{n_3(J_3+n_3)}} \,
\bar{\psi}_{J_1,n_1} \psi_{J_2,n_2} (\partial_z\partial_{\bar{z}}+z\bar{z}-1-
z\partial_z-\bar{z}\partial_{\bar{z}}) \, \psi_{J_3,n_{3-1}} \, .
\eq
Let's treat each term in $V_1$ separately.
We start from
\bea
T_1&\equiv& \int d^2 x \frac{1}{2\sqrt{\phi_0}\sqrt{n_3(J_3+n_3)}} \, \bar{\psi}_{J_1,n_1}
\psi_{J_2,n_2} \, \partial_z\partial_{\bar{z}} \, \psi_{J_3,n_{3-1}} \nn \\
&=& \int d^2 x \frac{1}{2\sqrt{\phi_0}\sqrt{n_3(J_3+n_3)}} \, \bar{\psi}_{J_1,n_1}
\psi_{J_2,n_2} \bigl[-l_0+z\bar{z}-1 \bigr] \psi_{J_3,n_{3-1}}
\nn \\
&=& \int d^2 x \frac{1}{2\sqrt{\phi_0}\sqrt{n_3(J_3+n_3)}} \, \bar{\psi}_{J_1,n_1}
\psi_{J_2,n_2} \bigl[-J_3-2n_3+z\bar{z}-1 \bigr]
\psi_{J_3,n_{3-1}}\; .
\ea
We then look at the term
\bea
T_2 &\equiv& -\int d^2 x \frac{1}{2\sqrt{\phi_0}\sqrt{n_3(J_3+n_3)}} \, \bar{\psi}_{J_1,n_1}
\psi_{J_2,n_2} \, z\, \pz \psi_{J_3,n_{3-1}} \nn \\
 &=& \int d^2 x \frac{1}{2\sqrt{n_3(J_3+n_3)}} \,
 \psi_{J_3,n_{3-1}} \psi_{J_2,n_2}
 \Bigl[\frac{1}{2\sqrt{\phi_0}} +\frac{z}{\sqrt{\phi_0}}\pz
 + \frac{z}{2}\pz \phi_0^{-1/2} \Bigr]\bar{\psi}_{J_1,n_1}+\nn \\
 &+& \int d^2 x \frac{1}{2\sqrt{n_3(J_3+n_3)}}\,
 \psi_{J_3,n_{3-1}}\bar{\psi}_{J_1,n_1}
 \Bigl[\frac{1}{2\sqrt{\phi_0}} +\frac{z}{\sqrt{\phi_0}}\pz
 + \frac{z}{2}\pz \phi_0^{-1/2} \Bigr] \psi_{J_2,n_2} \, .
\ea
Similarly,
\bea
T_3 &\equiv& -\int d^2 x \frac{1}{2\sqrt{\phi_0}\sqrt{n_3(J_3+n_3)}} \, \bar{\psi}_{J_1,n_1}
\psi_{J_2,n_2} \, \zb\, \pb \psi_{J_3,n_{3-1}} \nn \\
 &=& \int d^2 x \frac{1}{2\sqrt{n_3(J_3+n_3)}} \,
 \psi_{J_3,n_{3-1}} \psi_{J_2,n_2}
 \Bigl[\frac{1}{2\sqrt{\phi_0}} +\frac{\zb}{\sqrt{\phi_0}}\pb
 + \frac{\zb}{2}\pb \phi_0^{-1/2} \Bigr]\bar{\psi}_{J_1,n_1}+\nn \\
 &+& \int d^2 x \frac{1}{2\sqrt{n_3(J_3+n_3)}}\,
 \psi_{J_3,n_{3-1}}\bar{\psi}_{J_1,n_1}
 \Bigl[\frac{1}{2\sqrt{\phi_0}} +\frac{\zb}{\sqrt{\phi_0}}\pb
 + \frac{\zb}{2}\pb \phi_0^{-1/2} \Bigr] \psi_{J_2,n_2} \, .
\ea
We now collect all terms and, using the definitions of $l_0$ and
$l_-$, find
\bea
V_1 &=& \int d^2x
\frac{\psi_{J_3,n_{3-1}}}{\sqrt{\phi_0}\sqrt{n_3(J_3+n_3)}}
\;\;\;\; \times \nn \\
&\times&\Bigl[ \psi_{J_2,n_2} [l_- + \frac{l_0}{2} -z\zb+\half+ \frac{z\sqrt{\phi_0}}{4}\pz \phi_0^{-1/2}+
\frac{\zb \sqrt{\phi_0}}{4}\pb \phi_0^{-1/2}] \bar{\psi}_{J_1,n_1} + \nn \\
&+& \bar{\psi}_{J_1,n_1} [l_- + \frac{l_0}{2}-z\zb+\half
+ \frac{z\sqrt{\phi_0}}{4}\pz \phi_0^{-1/2}+ \frac{\zb \sqrt{\phi_0}}{4}\pb \phi_0^{-1/2}]  \psi_{J_2,n_2} +
\nn \\
&+& \bar{\psi}_{J_1,n_1} \psi_{J_2,n_2} \bigl[z\zb -1 -\frac{2n_3+J_3}{2}\bigr] \Bigr] \, .
\ea
We can rewrite the vertex above in the following way:
\bea
&&\int d^2x \frac{\bar{\psi}_{J_1,n_1}\psi_{J_2,n_2}}{\sqrt{\phi_0}\sqrt{n_3(J_3+n_3)}}
(l_+ +\frac{l_0}{2})\psi_{J_3,n_{3-1}}=\nn \\
&&= \int d^2x \frac{\psi_{J_3,n_{3-1}}\psi_{J_2,n_2}}{\sqrt{\phi_0}\sqrt{n_3(J_3+n_3)}}(l_-
+\frac{l_0}{2})\bar{\psi}_{J_1,n_1}+
\int d^2x \frac{\psi_{J_3,n_{3-1}}\bar{\psi}_{J_1,n_1}}{\sqrt{\phi_0}\sqrt{n_3(J_3+n_3)}}(l_-
+\frac{l_0}{2})\psi_{J_2,n_2}+\nn \\
&&+ \int d^2x \frac{\bar{\psi}_{J_1,n_1}\psi_{J_2,n_2}\psi_{J_3,n_{3-1}}}{\sqrt{\phi_0}\sqrt{n_3(J_3+n_3)}}
\Bigl[-z\zb  + \frac{z\sqrt{\phi_0}}{2}\pz \phi_0^{-1/2} +
\frac{\zb \sqrt{\phi_0}}{2}\pb \phi_0^{-1/2}  \Bigr]\, .
\ea
The last line vanishes, since $\phi_0=e^{-2z\zb}$.
Thus, we find the following identity:
\beq
\int d^2x
\frac{1}{\sqrt{\phi_0}}\Bigl[l_+^{(3)}-l_-^{(1)}-l_-^{(2)}+\frac{l_0^{(3)}-l_0^{(1)}-l_0^{(2)}}{2} \Bigr]
\bar{\psi}_{J_1,n_1}\psi_{J_2,n_2}\psi_{J_3,n_{3-1}}=0\, .
\eq
Notice that this identity can be used to relate the vertex for
single-matrix states (highest-weight states with $n_i=0$) to vertices of multi-matrix
states. In this sense, it provides a generating mechanism for
constructing non-trivial interactions starting from the (simpler) $1/2$ BPS sector
of the theory.
\subsection{Interactions in $AdS$}
We now move on to the case of real interest, interactions in $AdS
\times S$. For simplicity we consider $AdS_2 \times S^2$.
The generators are given by
\beq
l_{\pm}=i\left[\cos\rho \, \partial_\rho\mp i\sin\rho \, \partial_t\right],\qquad
l_0=i\partial_t \, ,
\eq
and the eigenfunctions (denoted by $\phi_n^\lambda$ to distinguish them from those of the
non-relativistic model) by
\beq
\phi_n^\lambda (t,\rho )=c(\lambda )
\sqrt{\frac{n!}{\Gamma (n+2\lambda)}}
\; e^{-i(n+\lambda)(t+{\frac{\pi}{2}})}
(\cos\rho )^\lambda \, C_n^\lambda (\sin\rho )\, ,
\eq
with
\beq
c(\lambda )={\Gamma (\lambda )\, 2^{\lambda -1}\over\sqrt{\pi}}\, .
\eq
Starting from $\int d\rho \big[ l_+^{(1)}\bar{\phi}_{n_1}^{\lambda_1}\big]\phi_{n_2}^{\lambda_2}\phi_{n_3}^{\lambda_3}$
at $t=0$ and integrating by parts with respect to $\rho$ we find
\bea
&& i\int d\rho \, \sin\rho \; \bar{\phi}_{n_1}^{\lambda_1}\phi_{n_2}^{\lambda_2}\phi_{n_3}^{\lambda_3}
+ i\int d\rho \, \sin\rho \big[l_0^{(2)}+l_0^{(3)}-l_0^{(1)}\big]\bar{\phi}_{n_1}^{\lambda_1}
\phi_{n_2}^{\lambda_2}\phi_{n_3}^{\lambda_3} \nn \\
&& +  \int d\rho \big[l_-^{(2)}+l_-^{(3)}-l_+^{(1)}\big]\bar{\phi}_{n_1}^{\lambda_1}\phi_{n_2}^{\lambda_2}\phi_{n_3}^{\lambda_3} =0
\ea
We can eliminate the $\sin\rho$ terms from the recursion relation by using
\beq
\sin\rho ={l_- -l_+\over 2il_0}
\eq
on the ``1" leg to obtain:
\bea
&& \int d\rho \left[ {l_-^{(1)}-l_+^{(1)}\over 2(n_1+\lambda_1) }\right]
\bar{\phi}_{n_1}^{\lambda_1}\phi_{n_2}^{\lambda_2}\phi_{n_3}^{\lambda_3} -\int d\rho \big[l_-^{(2)}+l_-^{(3)}-l_+^{(1)}\big]
\bar{\phi}_{n_1}^{\lambda_1}\phi_{n_2}^{\lambda_2}\phi_{n_3}^{\lambda_3}\nn \\
&& + \int d\rho \left[ {l_-^{(1)}-l_+^{(1)}\over 2(n_1+\lambda_1) }\right]\left[
n_1+\lambda_1+n_2+\lambda_2+n_3+\lambda_3\right]
\bar{\phi}_{n_1}^{\lambda_1}\phi_{n_2}^{\lambda_2}\phi_{n_3}^{\lambda_3}  =0.
\ea
\subsubsection{Use of the Ward Identity}
We can use this recursion relation to evaluate the overlap integral of a product
of \emph{any}
three eigenfunctions given the overlap of highest-weight eigenfunctions.
Inserting $n_1=n_1$, $n_2=n_3=0$ into the Ward identity and using ($C_0^\lambda = 1$,
$C_1^\lambda (x)=2\lambda x$)
\bea
&& l_- \, \phi_m^\lambda =e^{-it}\sqrt{m(m-1+2\lambda)}\, \phi_{m-1}^\lambda \, ,\qquad
l_- \, \bar{\phi}_n^\lambda = -e^{-it}\sqrt{(n+1)(n+2\lambda)} \,
\bar{\phi}_{n+1}^\lambda \, , \nn \\
&& l_+ \, \bar{\phi}_n^\lambda = -e^{+it}\sqrt{n(n-1+2\lambda)}\, \bar{\phi}_{n-1}^\lambda \, ,
\ea
we obtain
\bea
\label{WardOne}
&&\int d\rho \; \bar{\phi}_{n_1+1}^{\lambda_1}\phi_{0}^{\lambda_2}\phi_{0}^{\lambda_3} =
{1-n_1-\lambda_1+\lambda_2+\lambda_3\over 1+n_1+\lambda_1+\lambda_2+\lambda_3}
{\sqrt{n_1 (n_1-1+2\lambda_1
)}\over\sqrt{(n_1+1)(n_1+2\lambda_1)}}\;\times \nn \\
&&\times \;\; e^{2it}
 \int d\rho \; \bar{\phi}_{n_1-1}^{\lambda_1} \phi_{0}^{\lambda_2}\phi_{0}^{\lambda_3}.
\ea
This relation allows us to determine
$\int d\rho \; \bar{\phi}_{n_1}^{\lambda_1} \phi_{0}^{\lambda_2}\phi_{0}^{\lambda_3}$
for any $n_1$, once we know its value for $n_1=0,1$.
To obtain the value when $n_1=1$, insert $n_1=n_2=n_3=0$ into the Ward
identity.
The resulting identity implies that
$\int d\rho \; \bar{\phi}_{1}^{\lambda_1} \phi_{0}^{\lambda_2}\phi_{0}^{\lambda_3}=0$.
Next, set $n_1=n_1$, $n_2=n_2$ and $n_3=0$.
In this case, we find
\beq
\alpha_1\int d\rho \; \bar{\phi}_{n_1+1}^{\lambda_1}\phi_{n_2}^{\lambda_2}
\phi_{0}^{\lambda_3}+\alpha_2\int d\rho \; \bar{\phi}_{n_1-1}^{\lambda_1}
\phi_{n_2}^{\lambda_2}\phi_{0}^{\lambda_3}+\alpha_3\int d\rho \;
\bar{\phi}_{n_1}^{\lambda_1}\phi_{n_2-1}^{\lambda_2}\phi_{0}^{\lambda_3}=0,
\eq
where
\bea
\alpha_1 &=& -e^{-it}
\sqrt{(n_1+1)(n_1+2\lambda_1)}\;{1+n_1+\lambda_1+n_2+\lambda_2+\lambda_3\over
2(n_1+\lambda_1)} \, ,\\
\alpha_2 &=& e^{it}\sqrt{n_1(n_1-1+2\lambda_1)}\;{1-n_1-\lambda_1+n_2+\lambda_2+\lambda_3\over
2(n_1+\lambda_1)}\, ,\\
\alpha_3 &=& -e^{-it}\sqrt{n_2(n_2-1+2\lambda_2)} \, .
\ea
If we set $n_2=1$ we have
\beq
\alpha_1\int d\rho \; \bar{\phi}_{n_1+1}^{\lambda_1}\phi_{1}^{\lambda_2}\phi_{0}^{\lambda_3}
+\alpha_2\int d\rho \; \bar{\phi}_{n_1-1}^{\lambda_1}\phi_{1}^{\lambda_2}\phi_{0}^{\lambda_3}
+\alpha_3\int d\rho \; \bar{\phi}_{n_1}^{\lambda_1}\phi_{0}^{\lambda_2}\phi_{0}^{\lambda_3}=0,
\eq
which (starting from $n_1=0$) determines
$\int d\rho \; \bar{\phi}_{n_1}^{\lambda_1}\phi_{1}^{\lambda_2}\phi_{0}^{\lambda_3}$
for all $n_1$.
Next, set $n_2=2$ to obtain
\beq
\alpha_1\int d\rho \; \bar{\phi}_{n_1+1}^{\lambda_1}\phi_{2}^{\lambda_2}\phi_{0}^{\lambda_3}
+\alpha_2\int d\rho \; \bar{\phi}_{n_1-1}^{\lambda_1}\phi_{2}^{\lambda_2}\phi_{0}^{\lambda_3}
+\alpha_3\int d\rho \; \bar{\phi}_{n_1}^{\lambda_1}\phi_{1}^{\lambda_2}\phi_{0}^{\lambda_3}=0,
\eq
which (starting from $n_1=0$) fixes
$\int d\rho \; \bar{\phi}_{n_1}^{\lambda_1}\phi_{2}^{\lambda_2}\phi_{0}^{\lambda_3}$
for all $n_1$.
Continuing in this way, it is clear that we can determine
$\int d\rho \; \bar{\phi}_{n_1}^{\lambda_1}\phi_{n_2}^{\lambda_2}\phi_{0}^{\lambda_3}$,
for all $n_1,n_2$.
Finally, set $n_1=n_1$, $n_2=n_2$ and $n_3=n_3$. In this case, we find
\bea
&& \alpha_1\int d\rho \; \bar{\phi}_{n_1+1}^{\lambda_1}\phi_{n_2}^{\lambda_2}\phi_{n_3}^{\lambda_3}
+\alpha_2\int d\rho \; \bar{\phi}_{n_1-1}^{\lambda_1}\phi_{n_2}^{\lambda_2}\phi_{n_3}^{\lambda_3}
+\alpha_3\int
d\rho \; \bar{\phi}_{n_1}^{\lambda_1}\phi_{n_2-1}^{\lambda_2}\phi_{n_3}^{\lambda_3} \nn \\
&& +\alpha_4\int d\rho\bar{\phi}_{n_1}^{\lambda_1}\phi_{n_2}^{\lambda_2}\phi_{n_3-1}^{\lambda_3}=0,
\ea
where
\bea
\alpha_1 &=& -e^{-it}\sqrt{(n_1+1)(n_1+2\lambda_1)}\;{1+n_1+\lambda_1+n_2+\lambda_2+n_3+\lambda_3\over
2(n_1+\lambda_1)}\, ,\\
\alpha_2 &=& e^{it}\sqrt{n_1(n_1-1+2\lambda_1)}\;{1-n_1-\lambda_1+n_2+\lambda_2+n_3+\lambda_3\over
2(n_1+\lambda_1)}\, ,\\
\alpha_3 &=& -e^{-it}\sqrt{n_2(n_2-1+2\lambda_2)}\, , \\
\alpha_4 &=& -e^{-it}\sqrt{n_3(n_3-1+2\lambda_3)}\, .
\ea
If we take $n_2=0$ and $n_3=1$ we can determine
$\int d\rho \; \bar{\phi}_{n_1}^{\lambda_1}\phi_{0}^{\lambda_2}\phi_{1}^{\lambda_3}$
for all $n_1$.
Setting $n_2=1$ and $n_3=1$, we can find
$\int d\rho \; \bar{\phi}_{n_1}^{\lambda_1}\phi_{1}^{\lambda_2}\phi_{1}^{\lambda_3}$
for all $n_1$. Next, set $n_2=2$ and $n_3=1$ to get
$\int d\rho \; \bar{\phi}_{n_1}^{\lambda_1}\phi_{2}^{\lambda_2}\phi_{1}^{\lambda_3}$
for all $n_1$. Inching one step at a time we can determine the full vertex.
\subsubsection{Check of the Ward Identity}
To check the action of the generators we checked:
\bea
l_-\phi_0^\lambda &=& i\left[\cos\rho\partial_\rho +i\sin\rho \partial_t\right]\left(c(\lambda )
\sqrt{1\over \Gamma (2\lambda)}e^{-i\lambda(t+{\pi\over 2})}(\cos\rho )^\lambda\right)
=0\, , \\
l_-\bar{\phi}_1^\lambda &=& i\left[\cos\rho\partial_\rho + i\sin\rho \partial_t\right]
\left( c(\lambda )\sqrt{1\over \Gamma (1+2\lambda)}e^{i(1+\lambda)(t+{\pi\over 2})}
(\cos\rho )^\lambda 2\lambda \sin\rho\right) \nn \\
&=& -\sqrt{2(1+2\lambda )} \bar{\phi}_2^\lambda e^{-it}\, ,  \\
l_+\bar{\phi}_1^\lambda &=& i\left[\cos\rho\partial_\rho - i\sin\rho \partial_t\right]\left(
c(\lambda )\sqrt{1\over \Gamma (1+2\lambda)}e^{i(1+\lambda)(t+{\pi\over 2})}
(\cos\rho )^\lambda 2\lambda\sin\rho )\right)\nn \\
&=& -\sqrt{2\lambda}\bar{\phi}_0^\lambda e^{it}\, .
\ea
As a partial check of the results of the previous section, we will evaluate
(\ref{WardOne}) for $n_1=1$ and explicitly
verify that it is correct. After setting $n_1=1$ we have
\beq
\int d\rho \; \bar{\phi}_{2}^{\lambda_1}\phi_{0}^{\lambda_2}\phi_{0}^{\lambda_3} =
{\lambda_2+\lambda_3-\lambda_1\over 2+\lambda_1+\lambda_2+\lambda_3}
{\sqrt{2\lambda_1}\over\sqrt{2(1+2\lambda_1)}} \;\; e^{2it}
\int d\rho \bar{\phi}_{0}^{\lambda_1} \phi_{0}^{\lambda_2}\phi_{0}^{\lambda_3}.
\eq
Now,
\bea
&& \int d\rho \; \bar{\phi}_{2}^{\lambda_1}\phi_{0}^{\lambda_2}\phi_{0}^{\lambda_3} =
c(\lambda_1 )c(\lambda_2 )c(\lambda_3 )
\sqrt{2!\over \Gamma (2+2\lambda_1)\Gamma (2\lambda_2 )\Gamma
(2\lambda_3)}\;
e^{-i(\lambda_2-\lambda_3-\lambda_1)(t+\half\pi)}\nn \\
&& \times \; e^{2i(t+\half\pi)}\int d\rho \; (\cos\rho )^{\lambda_1+\lambda_2+\lambda_3}
\, C_2^{\lambda_1}(\sin\rho ).
\ea
Using
\bea
\sqrt{2!\over \Gamma (2+2\lambda_1)\Gamma (2\lambda_2 )\Gamma (2\lambda_3)}
&=&
\sqrt{2!\over (1+2\lambda_1)2\lambda_1}
\sqrt{1\over \Gamma (2\lambda_1)\Gamma (2\lambda_2 )\Gamma
(2\lambda_3)}\, ,\nn \\
e^{2i(t+\half\pi)}&=& -e^{2it}\, , \nn \\
\int d\rho \;(\cos\rho )^{\lambda_1+\lambda_2+\lambda_3}C_2^{\lambda_1}(\sin\rho )
&=&{\lambda_1 (\lambda_1-\lambda_2-\lambda_3)\over 2+\lambda_1+\lambda_2+\lambda_3}
\int d\rho \;(\cos\rho )^{\lambda_1+\lambda_2+\lambda_3}\, ,
\ea
it is trivial to verify the identity.

In conclusion, in this section we have demonstrated the existence of (SL(2))
Ward identities.
We have shown that these identities contain the necessary information
to specify the cubic interaction vertex
for general states from the knowledge of the vertex for the highest-weight states.
Since we have shown
that the one-matrix collective field theory correctly describes the latter case,
we therefore have a scheme of reconstruction of the full vertex.
This discussion was presented in the simplest $AdS_2$ framework;
it is clear, however, that this procedure is valid in general.
Nevertheless it will be important to develop the details in the higher dimensional case.
In particular, there should be significant
information on interactions in the 1/4 BPS sector where
progress has recently been accomplished
at the SUGRA level \cite{Donos:2006,Chen:2007du}.
\section{Acknowledgments}
One of us (AJ) would like to thank Tamiaki Yoneya for many discussions on the present
topic and the hospitality of the Theory Group at The University of Tokyo, Komaba
during part of this work. We also thank Aristomenis Donos for discussions.
The work of SC is supported in part by the Michigan
Society of Fellows. The work of RdMK is supported by the South African Research Chairs
Initiative of the Department of Science and Technology and National Research Foundation.

\appendix
\section{The Vertex in the two Representations}
\label{appendix}
In this appendix we show in some detail the origin
of the energy-conserving $\delta$-function in the 3-point function
in the original, \emph{non-tilde} representation.
Recall that a typical cubic term in the hamiltonian takes the form
\beq
\label{h3}
{H}_3^{(1)}= \frac{\omega_3(\omega_3-\omega_1-\omega_2)}{\sqrt{\omega_1\omega_2\omega_3}}
\, A_1 A_2 A_3^\dag \int d^D X \, \sqrt{-g}\,g^{tt} \;
\psi_1 \psi_2 \bar{\psi}_3 \, ,
\eq
where the integral is over the spatial coordinates only, and
the $AdS_{d+1}\times S^{d+1}$ wavefunction is given by
$\psi(x,y)=\sum_I f_I(x) Y^I(y)$.
Here $x$ and $y$ denote $AdS_{d+1}$ and $S^{d+1}$ coordinates respectively.
Furthermore, in general
we have $f(x)=f(t,\rho,\Omega_{d-1})=e^{-i\omega \,t}\Psi(\rho,\Omega_{d-1})$.
Incorporating the (trivial) time dependence $e^{-i\omega t}$ into the
creation/annihilation operators of (\ref{h3}), we see that the 3-vertex
\beq
V_3=
\int d^D X \sqrt{-g}\,g^{tt} \, \psi_1 \psi_2 \bar{\psi}_3
\eq
can be written as
\bea
V_3 &=&
\int d^{d}x \sqrt{-g_{AdS}}\,g^{tt}\,
\Psi_1 \Psi_2 \bar{\Psi}_3
\int d^{d+1}y \sqrt{g_{S}}\; Y_1 Y_2 \bar{Y}_3
\, , \equiv \, {\bf {\cal F}_{123}}\,{\bf {\cal G}_{123}}\, ,
\ea
where we defined
\beq
{\bf {\cal F}_{123}} \equiv \int d^{d}x \sqrt{-g_{AdS}}\,g^{tt}\, \Psi_1 \Psi_2 \bar{\Psi}_3,
\;\; \;\;\;{\bf {\cal G}_{123}} \equiv \int d^{d+1}y \sqrt{g_{S}}\; Y_1 Y_2 \bar{Y}_3 \, .
\eq
For simplicity, we now restrict ourselves to $AdS_2 \times S^2$, with
(global coordinates) metric
\beq
ds^2=-\sec{\rho}^2dt^2+\sec{\rho}^2d\rho^2+\sin{\theta}^2 d\phi^2
+ d\theta^2\, .
\eq
On the sphere one has
\beq
\Box_{S^2}Y_j^{\bar{m}}=-j(j+1)Y_j^{\bar{m}},
\eq
where $Y_j^{\bar{m}}(\phi,\theta)=\tilde{N}_j^{\bar{m}} \,
e^{i{\bar{m}}\phi}P_j^{\bar{m}}(\cos{\theta})$,
$\bar{m}=-j,-j+1,\ldots,j$ and $\tilde{N}_j^{\bar{m}}$ is the
proper normalization.
The AdS wavefunctions which satisfy the wave equation
\beq
\Box_{AdS_2}f=\cos{\rho}^2(-\p_t^2+\p_\rho^2)f=m^2 f
\eq
are given by
\beq
f_{\omega,\lambda}(t,\rho)=N_{\omega-\lambda}^\lambda
e^{-i\omega \,t}\, (\cos{\rho})^\lambda \, C_{\omega-\lambda}^\lambda(\sin{\rho}), \;\;\;\;
\omega=\lambda+n, \;\;\;n=0,1,2,\ldots \; .
\eq
Here $C_{\omega-\lambda}^\lambda(\sin{\rho})$ are Gegenbauer
polynomials, $N_{\omega-\lambda}^\lambda$ is a normalization
factor and $\lambda$ is related to the mass of the field
(also note $0 \leq \rho < \frac{\pi}{2}$).
For chiral primaries the mass turns out to be
$m^2=j(j-1)$,
and the highest weight state is given by $\lambda=j$ and $n=0$:
\beq
\psi_{h.w.}(t,\rho,\phi,\theta)=N_0^{j}\,\tilde{N}_j^j \; e^{-ij \,t} \,(\cos{\rho})^j \,
e^{ij\phi} P_j^j(\cos{\theta}).
\eq
In order for the spherical harmonics and the Gegenbauer polynomials to be
$\delta$-function normalized we must take
\beq
\tilde{N}_j^{\bar{m}} = \sqrt{\frac{(2j+1)(j-\bar{m})!}{4\pi
(j+\bar{m})!}}\, , \;\;\;\;
N_n^\lambda=\frac{\Gamma(\lambda)2^{\lambda-1/2}}{\sqrt{\pi}}\sqrt{\frac{n!
(n+\lambda)}{\Gamma(n+2\lambda)}}\, .
\eq
For highest-weight states on $AdS_2 \times S^2$ the overlap integrals are
(defining $j\equiv j_1+j_2+j_3$):
\bea
{\bf {\cal F}}_{j_1 j_2 j_3}&=&\int_0^{\frac{\pi}{2}} d\rho \sqrt{-g_{AdS_2}}\,g^{tt}\,
\Psi_{j_1} \Psi_{j_2} \bar{\Psi}_{j_3} \nonumber \\
&=& N_0^{j_1} N_0^{j_2} N_0^{j_3}\int d\rho (\cos{\rho})^j \, C_0^{j_1} C_0^{j_2}
C_0^{j_3} \nonumber \\
&=&N_0^{j_1} N_0^{j_2} N_0^{j_3} \Biggl( \frac{\sqrt{\pi}}{2}
\frac{\Gamma(\frac{1}{2}+\frac{j}{2})}{\Gamma(1+\frac{j}{2})}
\Biggr)\, , \\
{\bf {\cal G}}_{j_1 j_2 j_3}&=&\int d^{2}y \sqrt{g_{S^2}}\; Y_{j_1}^{j_1}
Y_{j_2}^{j_2}\bar{Y}_{j_3}^{j_3} \nonumber \\
&=& \delta(j_1+j_2-j_3) \, \frac{\tilde{N}_{j_1}^{j_1} \tilde{N}_{j_2}^{j_2}
\tilde{N}_{j_3}^{j_3}}
{(2\pi)^{3/2}}\int_0^\pi d\theta
\,\sin{\theta}\, P_{j_1}^{j_1}P_{j_2}^{j_2}P_{j_3}^{j_3}\nonumber \\
&=& \delta(j_1+j_2-j_3) \frac{(-1)^j\sqrt{\pi}}{(2\pi)^{3/2}}
\prod_{i=1}^3 \Bigl((2j_i-1)!! \; \tilde{N}_{j_i}^{j_i} \Bigr)
\frac{\Gamma(1+\frac{j}{2})}{\Gamma(\frac{3}{2}+\frac{j}{2})},
\ea
where we used $C_0^j=1$ and $P_j^j(x)=(-1)^j (2j-1)!! (1-x^2)^{j/2}$.
Note the appearance of the $\delta(j_1+j_2-j_3)$ term in ${\bf {\cal G}}_{j_1 j_2 j_3}$,
coming from the $\int d\phi \; e^{ij(\phi_1+\phi_2-\phi_3)}$ integral.
After some Gamma function cancellations, we are left
with the following 3-vertex:
\bea
{\bf {\cal F}}_{j_1 j_2 j_3} \, {\bf {\cal G}}_{j_1 j_2 j_3}
&=& \delta(j_1+j_2-j_3) \, \frac{(-1)^j \, \pi}{(2\pi)^{3/2}} \, \,
\prod_{i=1}^3 \Bigl((2j_i-1)!! N_0^{j_i}\, \tilde{N}_{j_i}^{j_i} \Bigr)
\frac{1}{j+1},
\ea
where we used $\Gamma(n/2)=\sqrt{2\pi}\,(n-2)!! \,2^{-n/2}$.
Finally, using
\beq
N_0^{j_i}=\frac{\Gamma(j_i)\, 2^{j_i}\,\sqrt{j_i}}{\sqrt{2\pi \, \Gamma(2j_i)}
}\, , \;\;\;\;\;\;\; \tilde{N}_{j_i}^{j_i}=\sqrt{\frac{2j_i+1}{2\,
(2j_i)!}}\, , \;\;\; \text{and} \;\;\; (2j_i-1)!!=\frac{(2j_i)!}{2^{j_i}
\,(j_i)!} \, ,
\eq
we find
\beq
(2j_i-1)!! N_0^{j_i}\, \tilde{N}_{j_i}^{j_i} =
\sqrt{\frac{2j_i+1}{2\pi}}.
\eq
The final expression for the 3-point function overlap is
\beq
V_3 \equiv {\bf {\cal F}}_{j_1 j_2 j_3} \, {\bf {\cal G}}_{j_1 j_2 j_3}
= \delta(j_1+j_2-j_3)  \,
\Biggl[ \frac{(-1)^j}{(8\pi^2)(j+1)} \sqrt{(2j_1+1)(2j_2+1)(2j_3+1)}\Biggr] .
\eq
Thus, (for the simple case of $AdS_2 \times S^2$) we have explicitly shown the origin
of the delta-function term, and presented the final expression for the 3-point
overlap.
This calculation can be repeated for the more general wavefunctions
given in (\ref{LagWavefn}). The 3-point overlap will then be
roughly of the form
\beq
V_3 \sim \int d\phi \; e^{i(j_1+j_2-j_3)\phi} \; \int dr \, L_{n_1}^{j^1} \,
L_{n_2}^{j^2}\, L_{n_3}^{j^3} \sim
\delta(j_1+j_2-j_3) \int dr \, L_{n_1}^{j^1} \,
L_{n_2}^{j^2}\, L_{n_3}^{j^3} \, ,
\eq
and still yield a conserving delta
function.
On the other hand, in the case of the \emph{tilde} representation
the wavefunctions are
\beq
\tilde{\psi}(x,y) = e^{-(x^2+y^2)}H_{j+n}(x) \, H_n(y) \, ,
\eq
and the overlap integral takes the form
\beq
V_3 \sim \int dx  \; e^{-3x^2} H_{j_1+n_1}(x) \, H_{j_2+n_2}(x) \,
H_{j_3+n_3}(x)
\int dy \; e^{-3y^2}H_{n_1}(y) \, H_{n_2}(y) \, H_{n_3}(y)\, .
\eq
As one can verify, written in this basis the vertex no longer has a conserving
$\delta$-function.

\end{document}